\documentclass{ws-ijmpd}

\usepackage{graphicx}
\usepackage{bm}
\usepackage{amssymb}

\newcommand{\be}{ \begin{equation}}
\newcommand{\bea}{ \begin{eqnarray}}
\newcommand{\eea}{ \end{eqnarray}}
\newcommand{\ee}{\end{equation}} 
\newtheorem{defi}{Definition}

\begin{document}

\markboth{Thomas P Sotiriou, Valerio Faraoni and Stefano 
Liberati}
{Theory of gravitation theories: a no-progress report}

%
\catchline{}{}{}{}{}
%

\title{THEORY OF GRAVITATION THEORIES: A NO-PROGRESS REPORT}

\author{Thomas P Sotiriou$^1$, Valerio Faraoni$^2$, Stefano 
Liberati$^1$}
\address{$^1$ SISSA - International School for Advanced Studies, Via Beirut 2-4, 34014, Trieste, Italy and 
INFN, Sezione di Trieste\\ sotiriou@sissa.it, liberati@sissa.it}
\address{$^2$ Physics Department, Bishop's University, 
Sherbrooke, Qu\'{e}bec, Canada J1M 1Z7\\ vfaraoni@ubishops.ca}





\maketitle

\begin{history}
\received{Day Month Year}
\revised{Day Month Year}
\comby{Managing Editor}
\end{history}

\begin{abstract}
Already in the 1970s there where attempts to 
present a set of ground rules, sometimes referred to as a theory 
of gravitation theories, which theories of gravity should 
satisfy in order to be considered  viable in principle and, 
therefore, interesting enough to deserve further investigation. 
From this perspective, an alternative title of the present paper 
could  be ``why are we still unable to write a guide on how to 
propose 
viable alternatives to general relativity?''. Attempting to  
answer this question, it is argued here that earlier efforts to 
turn qualitative statements, such as the Einstein 
Equivalence Principle, into quantitative ones, such as the metric 
postulates, stand on rather shaky grounds --- probably contrary 
to popular belief --- as they appear to depend strongly on 
particular representations of the theory. This includes 
ambiguities in the identification of matter and gravitational 
fields, dependence of frequently used definitions, such as 
those of the stress-energy tensor or 
classical vacuum, on the choice of variables, {\em etc}. Various 
examples are discussed and possible approaches to this problem 
are pointed out. In the course of this study, several common 
misconceptions related to the various forms of the Equivalence 
Principle, the use of conformal frames and equivalence between 
theories are clarified. 
\end{abstract}


\section{Introduction} \label{intro} 

The axiomatic 
formulation of general relativity (or gravitational theories in 
general) seems to resemble the  myth of the Holy Grail. Serious 
attempts have been made to find it and everybody seems to be 
interested in it, but nobody actually knows where to look for it. 
Of course, one could ask how useful a collection of axioms 
could be for a theory, like general relativity, for which we already 
know the field equations or the action. Indeed, knowledge of any 
of the latter suffices to fully describe the dynamics of the 
theory, at least at the classical level, hence making the absence 
of an axiomatic formulation less needed, at least for practical 
purposes. 

 Nonetheless, there are numerous reasons why one would like to explore the 
conceptual basis of a theory. On the theoretical side, one 
could mention that a set of axioms could help us understand 
the theory in depth and provide a better insight in finding 
solutions to long standing problems, the most prominent of which being to propose a theory of quantum gravity (for example it could help in determining the fundamental classical properties one should expect to recover and recognize which of them could break down at the quantum level).

Furthermore, if emergent gravity scenarios are considered ({\em 
i.e.,}~scenarios in which the metric and the affine-connection 
are collective variables and general relativity  would be a sort 
of  hydrodynamics emergent from more fundamental constituents) 
such  a set of axioms could provide a much needed guidance in 
reconstructing  the microscopic system at the origin of 
classical gravitation, for  example by constraining its 
microscopic properties so to  reproduce the emergent physical 
features encoded in these axioms.

There could be important benefits at the purely  experimental level as well. Past experience 
taught us that experiments test principles and not theories 
 (for example weak equivalence principle tests, such as the 
gravitational shift ones\cite{PoundRebka}, were initially 
erroneously regarded as tests of general relativity). So one would want 
to know exactly which principles/axioms to test in order to discriminate at least among classes of gravitational theories.

Finally, nowadays we have a number of alternative theories 
of gravity. How can we  characterise the way in which they differ from general relativity, 
group them, or obtain  some insight into which of them are 
preferable with respect to others? 
Even if we are far from a coherent and strict axiomatic 
 formulation, at least a set of principles, or what is sometimes 
called a theory of gravitation theories ({\em i.e.,}~a 
``meta--theory of gravitation"), 
would definitely prove useful to this end, allowing us not only 
to ``catalogue" presently known theories but also to build new 
ones by just relaxing or adding some fundamental 
principles/axioms.
Last but not least, having such a meta--theory would come up 
extremely handy  given, for example, the current revival of 
alternative theories of gravitation  as a possible explanation 
of  nowadays puzzling cosmological observations ({\em e.g.,}~the 
dark 
energy problem). Indeed, having such a theory would greatly help 
in interpreting the experimental 
results and their  implications for discriminating among  
alternative theories of gravitation.

Considering all of the above, it is quite disappointing 
that no real progress has been made in this direction in  the 
last thirty years. Indeed, it  seems that the scientific community has somewhat given up on such an ambitious task given that the latest serious attempts date back to  the 70's, even thought the subject of alternative theories of gravity has been quite an active one throughout this time. We feel that it is important to understand 
the practical reasons for this lack of progress if we wish to step forward in this research and go beyond the trial-and-error approach that is mostly used in modified gravity. 
Hopefully, this exploration will also give, as a byproduct, interesting clarifications relevant to certain   common  misconceptions (weak equivalence principle, equivalence of 
theories, {\em etc.}) and  maybe even serve as a motivational point of reference for future work.

In Sections \ref{equivprinc} and \ref{metricpost} we formulate  
the problem in general terms by analysing the several formulations of the Equivalence Principle and their implementations via the so called metric postulates.   In Section~\ref{repres} we 
distinguish between theories and their representations. Concrete 
examples are given in the following sections, 
including scalar-tensor theories\cite{VFbook} in Section \ref{ST}, the metric and Palatini 
versions of $f(R)$  gravity\cite{buh,thesis,fer,fofrcosm,viability} in Section \ref{fR}, 
and the  Einstein-Cartan-Sciama-Kibble theory\cite{hehl} in Section \ref{ECSK}.   
Finally, Section \ref{discussion} contains a discussion of the results  
and our conclusions.

In what follows purely classical physics will be considered. The 
issue of the compatibility between the Equivalence Principle(s) and 
quantum mechanics, although rich in facets and consequences\cite{EPQM} is beyond the scope of this work.
The  metric is taken to have signature $(-,+,+,+)$, we define 
$ \kappa=8\,\pi\,G$ where $G$ is Newton's constant and $c$ the 
speed of light in vacuum,  units where $G=c=1$ are used, and the basic notation of Wald's book\cite{Wald} is adopted.

\section{Equivalence Principle(s)}
\label{equivprinc}

 In creating a meta--theory of gravitation one immediately 
faces the daunting task to provide some sufficiently general 
 criteria for characterizing the physical features of possible 
 alternative theories. As already mentioned, providing a strict axiomatic 
formulation is hardly an easy goal~\footnote{There has been, however, an attempt towards an axiomatic 
formulation of gravitational theories from a more 
mathematically-minded point of view\cite{mathapproach}.}, but one could hope to give at least some set of physical viability principles, even if the latter are not necessarilly at the level of axioms. It is clear that in order to be useful 
 such statements need to be formulated in a theory-independent 
 way and should be amenable to experimental tests so that we 
could  select at least among classes of gravitational theories 
by suitable  observations/experiments. The best example in this 
direction has  been so far the Equivalence Principle in its 
various versions\cite{Willbook}.  Let us restate them  here  and 
then pass to analyze  their physical implications:

{\it Weak Equivalence Principle (WEP)}: If an uncharged test 
body is placed at an initial event in spacetime and given an initial 
velocity there, then its subsequent trajectory will be 
independent of its internal structure and composition.

{\it Einstein Equivalence Principle (EEP)}: (i) WEP is valid, 
(ii) the 
outcome of any local non-gravitational test experiment is 
independent of the velocity of the freely falling apparatus 
(Local Lorentz Invariance or LLI) and (iii) the outcome of any 
local non-gravitational test experiment is independent of where 
and when in the universe it is performed (Local Position 
Invariance or LPI).

{\it Strong Equivalence Principle (SEP)}: (i) WEP is valid for 
self-gravitating bodies as well as for test bodies, (ii) the 
outcome of any local test experiment is independent of the 
velocity of the freely falling apparatus (Local Lorentz 
Invariance or LLI) and (iii) the outcome of any local test 
experiment is independent of where and when in the universe it 
is performed (Local Position Invariance or LPI).

It is important to stress that the WEP only says that there 
exist some preferred trajectories, the free fall trajectories, 
that test particles will follow and  these curves are the same 
independently of the mass and internal 
composition of the particles that follow them (universality of  
free fall). WEP does not imply, by itself, that there 
exist a metric, geodesics, {\em etc.} --- this comes about only 
through the EEP by combining the WEP with requirements (ii) and 
(iii) (see Will's book\cite{Willbook} for a detailed discussion). The 
same is 
true for  the covariance of the field equations. As far as the SEP is 
concerned, the main thrusts  extending the validity of the WEP to 
self-gravitating bodies and the applicability of LLI and LPI to gravitational experiments, in contrast to the EEP.
There are experimental tests of all the EPs but the stringent 
ones are for the WEP and the EEP. 

 Let us stress that there are at least three subtle points in 
 relation to the use and meaning of the EP formulations, the 
 first one concerning the relation between the SEP and general 
relativity.  While there are claims that SEP holds only for general relativity\cite{Willbook}, no proof of this statement has been given 
so far. Indeed, it would be a crucial step forward to  pinpoint 
a one-to-one association between GR and the SEP  but it is difficult to relate directly and  
uniquely a qualitative statement, such as the SEP, to a  
quantitative one, namely Einstein's equations.
 The second subtle point is the  reference to test particles in 
all the EP formulations. Apart from the obvious limitation of restricting attention to  particles and ignoring classical fields (such as, e.g., the  
 electromagnetic one), apparently no true test 
particles exist, hence the question is how do we know how 
``small'' a particle should be in order to be considered a test 
particle ({\em i.e.}, its gravitational field can be neglected)? 
The  answer is likely to be theory-dependent (see {\em e.g.}, Geroch and Jang\cite{Geroch} and references therein for the  case of general relativity), so there is no 
guarantee that a theory cannot be put together in which the WEP is 
valid in principle but, in practice, experiments would show a 
violation because, within the framework 
of the theory, a ``small'' particle is not close enough to a test 
particle. Of course, such a theory would not be viable but 
this would not be obvious when we refer only to the WEP 
from a theoretical perspective ({\em e.g.}, calculate free fall 
trajectories and compare with geodesics). A third subtlety, 
on which we shall come back later, is related to the fact that 
sometimes the same theory appears to evidently satisfy or not 
some version of the EP depending on which variables are used in 
describing it, an example being the Jordan versus the Einstein 
frame in scalar-tensor theories of gravity.

Taking all of the above into consideration, it seems that the main problem with all forms of the equivalence 
principle is that they are of little practical value. As 
principles they are by definition qualitative and not 
quantitative. However, quantitative statements are  what is 
needed in practice. An attempt to overcome this difficulty was 
indeed made by  Thorne and Will\cite{ThorneWill} and is   
embodied by the so called metric theories postulates.

\section{Metric  Postulates} 
\label{metricpost}

The  metric postulates can be stated in the following 
way\cite{ThorneWill}:

\begin{enumerate}

\item there exists a metric $g_{\mu\nu}$ (second rank non 
degenerate tensor).

\item $\nabla_{\mu}T^{\mu\nu}=0$, where $\nabla_{\mu}$ is the 
covariant derivative defined with the Levi-Civita connection of 
this metric and $T_{\mu\nu}$ is the stress-energy tensor of 
non-gravitational (matter) fields. 
\end{enumerate}

Theories that satisfy the metric postulates are often called 
{\em metric theories}. Let us first see how these postulates 
come about starting from the EEP and how they encapsulate its 
validity. 
The EEP adds two more statements to the WEP: Local Lorentz 
Invariance and Local Position invariance. A freely falling 
observer carries a local frame in which test bodies have 
unaccelerated motions. According to the requirements of LLI the 
outcomes of non-gravitational experiments are independent of the 
velocity of the freely falling frame and therefore, if two such 
frames located at the same event ${\cal P}$ have different 
velocities, this should not affect the predictions of identical 
non-gravitational experiments. Local Position Invariance 
requires that the above should hold for all spacetime points. 
Therefore, roughly speaking, in local freely falling frames the 
theory should reduce to special relativity.

This implies that there should be (at least one) second rank 
tensor 
fields which reduce, in the local freely falling frame, to metrics 
conformal to the Minkowski one. The freedom of having an 
arbitrary conformal factor is due to the fact that the EEP does 
not forbid a conformal rescaling in order to arrive to 
special-relativistic expressions of the physical laws in the 
local freely falling frame. Note, however, that while one could 
think to allow each specific matter field to be coupled to a different one of these conformally related second rank tensors, the conformal factors relating these 
tensors can at most differ by a multiplication constant if the couplings to different matter fields are to be turned to constants under a conformal rescaling as the LPI 
requires (this highlights the relation between LPI and varying 
coupling constants). We can then conclude that rescaling 
coupling constants and performing a conformal transformation 
leads to a metric $g_{\mu\nu}$ which, in every freely falling  
local frame, reduces (locally) to the Minkowski metric  
$\eta_{\mu\nu}$ \footnote{This does not exclude the possibility 
of having a second metric tensor in the theory  as long as this 
metric does not couple to the matter (this case leading to theories of the  
bi-metric kind).}.

It should be stressed that all conformal metrics $\phi 
g_{\mu\nu}$, $\phi$ being the conformal factor, can be used to 
write down the equations or the action of the theory. 
$g_{\mu\nu}$ is only special in the following sense: Since at 
each event ${\cal P}$ there exist local frames called local 
Lorentz frames, one can find suitable coordinates in which at 
${\cal P}$ 
\be \label{localmet} 
g_{\mu\nu}=\eta_{\mu\nu}+{\cal 
O}\left(\sum_\alpha |x^\alpha-x^\alpha({\cal P})|^2\right), 
\ee 
and $\partial g_{\mu\nu}/\partial x^\alpha=0$.  In local Lorentz 
frames the geodesics of the metric $g_{\mu\nu}$ are straight 
lines. Free fall trajectories are straight lines in a local 
freely falling frame. Identifying the two frames we realize that 
the geodesics of $g_{\mu\nu}$ coincide with free fall 
trajectories. Put in other words, the EEP requires the existence 
of a family of conformal metrics, one of which should have 
geodesics which coincide with free fall trajectories. On the 
other hand, geodesic motion for test particles can be derived 
from the condition $\nabla_{\mu}T^{\mu\nu}=0$, when $\nabla_\mu$ 
is the covariant derivative defined by the Levi-Civita 
connection of the metric, whose geodesics the test particles 
have to follow\cite{fock}.

Appealing as they may seem, however, the metric postulates lack 
clarity. As pointed out also by the very authors of 
the paper introducing them\cite{ThorneWill}, any metric theory can perfectly well be 
given a representation that appears to violate the metric postulates (recall, for instance, that $g_{\mu\nu}$ is a member of a family of conformal 
metrics and that there is no {\it a priori} reason why this 
metric  should be used to write down the field equations). 
 See also Anderson\cite{Anderson} for an earlier criticism 
of the need for a metric and, indirectly, of the metric 
postulates. One 
of the goals of this paper is to elaborate on the problems mentioned above, 
as well as on other prominent ambiguities stated below and trace 
their roots. 

\subsection{What is precisely the definition of  stress-energy 
tensor?}
\label{set}

In order to answer this question one could refer to an action. 
This is a significant restriction to begin with, since it would 
add to the 
EEP the prerequisite that a reasonable theory has to come from 
an action. Even so, such an assumption would not solve the problem: one 
could claim that $T_{\mu\nu}\equiv -(2/\sqrt{-g})\delta S_M/\delta 
g^{\mu\nu}$ but then how is the matter action $S_M$ defined? 
Claiming that it is the 
action from which the field equations for matter are derived  is 
not sufficient since it does not provide any insight about the 
presence of the gravitational fields in $S_M$. Invoking a 
minimal 
coupling argument, on the other hand, is strongly theory-dependent (which coupling 
is really minimal in a theory with extra fields or, say, an 
independent connection? \cite{metaff1}). Furthermore, whether a matter field couples minimally or 
non-minimally to gravity or to matter should be decided by 
experiments. Since a non-minimal coupling could be present and 
evade experimental detection (as proposed in string theories\cite{TaylorVeneziano}), it seems prudent to allow for it in the 
action or the theory.

Setting actions aside and resorting to the correspondence with 
the stress-energy tensor of special relativity does not 
help either. There is always more than one tensor that one can 
construct, which will reduce to the special-relativistic 
stress-energy tensor when gravity is ``switched off''. Moreover
it is not clear what ``switched off'' exactly means when extra  
fields describing gravity (scalar or vector) are present in the 
theory together with the metric tensor.

Finally, mixing the two tentative  definitions described above 
makes the situation even worse: one can 
easily imagine theories in which 
$T_{\mu\nu}=-(2/\sqrt{-g})\delta S_M/\delta g^{\mu\nu}$ does not 
reduce to the special-relativistic stress-energy tensor in some 
limit. Are these theories necessarily non-metric? This point 
highlights also another important question: are the metric 
postulates a  necessary or a sufficient condition for the 
validity of the EEP? Concrete examples are provided in sections 5, 6 and 7.

\subsection{What does ``non-gravitational fields'' mean?}
\label{matgeom}
There is no precise definition of ``gravitational'' and 
``non-gravitational'' field. One could say that a 
field non-minimally coupled to the metric is gravitational 
whereas the rest are matter fields. This definition does not 
appear to be rigorous or sufficient and  it is 
shown in the following that it strongly depends on the 
perspective and the terminology one chooses.

Consider, for example, a scalar field $\phi$ 
non-minimally coupled to the 
Ricci curvature in $\lambda \phi^4 $ theory, as described by 
the action
\begin{equation} \label{NMCaction}
S=\int d^4x \, \sqrt{-g}\, \left[ \left( \frac{1}{2\kappa}-\xi 
\phi^2 \right) R-\frac{1}{2}\nabla^{\mu}\phi\nabla_{\mu}\phi  
-\lambda \phi^4 \right] \;.
\end{equation}
If one begins with  a classical scalar field minimally coupled 
to the curvature ({\em i.e.}, $\xi=0$) in the potential 
$\lambda \phi^4 $ and quantizes it, one finds that first 
loop corrections  prescribe a non-minimal coupling term 
({\em i.e.}, 
$\xi\neq 0$) if the 
theory is to be renormalizable, thus 
obtaining the ``improved energy-momentum tensor'' of Callan, 
Coleman, and Jackiw\cite{CCJ} (see also 
Chernikov and Tagirov\cite{ChernikovTagirov}). Does quantization change the 
character of this scalar field from ``non-gravitational'' to 
``gravitational''? Formally, the resulting theory is a 
scalar-tensor theory according to every definition of such 
theories that one finds in the literature\cite{VFbook,Willbook,Wagoner,FujiMaeda}, but many authors 
consider $\phi$  a non-gravitational field, and certainly this 
is the point of view of Callan {\em et al}\cite{CCJ} (in 
which $\phi$ is regarded as a matter field  being quantized) and 
of most
particle physicists.

\section{Theories and representations}
\label{repres}

As it will be demonstrated later on, many misconceptions arise 
when a theory is identified with one of its  representations and 
other 
representations are implicitly treated as different theories. 
Even though this might seem too abstract, to 
avoid confusion, one would like to provide precise definitions of 
the words ``theory'' and ``representation''. 
This, however, is not trivial. For the term ``theory'', even if 
one looks at a popular internet dictionary, a number of possible   
definitions can be found\cite{wiktion}: \begin{enumerate} 

\item An unproven conjecture. 
\item An expectation of what should happen, barring unforeseen 
circumstances. 

\item A coherent statement or set of statements 
that attempts to explain observed phenomena. 

\item A logical structure that enables one to deduce the 
possible results of every experiment that falls within its 
purview. 

\item A field of  study attempting to exhaustively describe a 
particular class of constructs. 

\item A set of axioms together with all statements 
derivable from them. 
\end{enumerate}

It is apparent that definitions (i) and (ii) are not applicable 
to physical theories. On the other hand, (iii) and (iv) seem to 
be complementary statements describing the use of the word 
``theory'' 
in natural sciences, whereas (v) and (vi) appear to have a 
mathematical and logical basis respectively. In a loose sense, a 
more complete definition for the word ``theory'' in the context 
of 
physics would probably come from a combination of (iv) and (vi), 
in order to combine the reference to experiments in (iv) and the 
mathematical rigidity of (vi). An attempt towards this direction 
could be: 

\begin{defi} Physical Theory: A coherent logical structure, 
preferably expressed through a set of axioms together with all 
statements derivable from them, plus a set of rules for 
their physical interpretation, that enable one to deduce and 
interpret the 
possible results of every experiment that falls within its 
purview.\footnote{ One might argue that when a theory is defined as a set of axioms, as suggested above, it is doomed to face the implications of G\"odel's incompleteness theorems. However, it is neither clear if  such theorems are applicable to physical theories, nor how  physically relevant they would be even if they were applicable\cite{Godel,Barrow:2006hi}.}
\end{defi}

Note that no reference is made as to whether there is an 
agreement between the predictions of the theory and the actual 
experiments \footnote{It is, however, necessary that a theory 
makes some  predictions which are testable in principle before 
being called a physical theory.}. This is a further step which should be 
considered in order to check how successful a theory is in describing the physical world.  There could be criteria 
according to which  the theory is successful or not according to 
how large a class of observations is explained by it and to
the level of accuracy obtained (see for example Hawking's book\cite{hawk}). 
Additionally, one could consider simplicity as a merit and 
characterize a theory according to the 
number of assumptions that it is based on (Ockham's razor). 
However, all the above should not be included in the definition 
of the word ``theory'' itself.

Physical theories should acquire a mathematical representation. 
This requires the introduction of physical variables 
(functions or fields) with which the axioms can be encoded in 
mathematical relations. We attempt to give a definition:

\begin{defi} Representation (of a theory): A finite  
collection of equations interrelating the physical variables 
which are used to describe the elements of a theory and 
assimilate its axioms. 
\end{defi}

The reference to equations can be restrictive as one may claim 
that in many cases a theory could be represented fully by an 
action. At the same time it is obvious that a representation of 
a theory is far from being unique. Therefore, one might prefer 
to 
modify the above definition as follows:

\begin{defi} Representation (of a theory): A non-unique choice 
of physical variables between which, in a prescribed way, one 
can form inter-relational expressions that assimilate the axioms 
of the theory and can be used in order to deduce derivable 
statements. \end{defi}

It is worth stressing here that when choosing a representation for a theory it is 
essential to provide also a set of rules for the physical interpretation of the variables involved in it. This is needed for formulating the axioms ({\em i.e.}, the physical statements) of the theory in terms of these variables.  It should also be noted that these rules come as extra information not {\em a priori} contained in the mathematical formalism. Furthermore, once they are consistently used to interpret the variables of the latter, they would allow to consistently predict the outcome of experiments in any alternative representation (we shall come back to this point and discuss an example later on in section \ref{vacuum}).

All of the above definitions are, of course, tentative or even naive ones and others can be found that are more precise and comprehensive. However, they are good enough to make  the following point: the arbitrariness that inevitably exists in 
choosing the physical variables is bound to affect the 
representation. More specifically, it will affect the clarity 
with which the axioms or principles of the theory appear in 
every  representation. Therefore, there will be representations in 
which it will be obvious that a certain principle is satisfied 
and others in which it will be more intricate to see that. 
However, it is clear that the theory is one and the same and 
that the  axioms or principles are independent of the representation. 


\section{Scalar-tensor gravity}
\label{ST}

In order to make the discussion of the previous sections clearer, 
let us use scalar-tensor theories of gravitation as an example. 
As in most current theories, scalar-tensor theory was not 
originally introduced as a collection of axioms but directly 
through a representation. More precisely, this class of theories 
is described by the action 
\begin{equation} 
S=S^{(g)}+S^{(m)}\left[e^{2\alpha(\phi)}g_{\mu\nu}, 
\psi^{(m)}\right]\;,\label{eq:3.19}
\end{equation} 
where 
\begin{equation} \label{action} 
S^{(g)}=\int 
d^{4}x\;\sqrt{-g}\left[\frac{A(\phi)}{16\pi 
G}R- \frac{B(\phi)}{2}g^{\mu\nu}\nabla_{\mu} 
\phi\nabla_{\nu}\phi-V(\phi)\right] \;.
\end{equation}
In order to write this action we have used the notation of 
Flanagan\cite{Flanagan} (see also Shapiro and Takata\cite{shapiro}). Note that this is not the most 
conventional 
notation found in the literature, as some of the unspecified 
functions $A$, $B$, $V$, and $\alpha$ can be fixed without loss 
of generality, {\em i.e.,~}without choosing a theory within the 
class. However, this would come at  the expense of fixing 
the  representation, which is exactly what we intend to analyse 
here.  Therefore, this notation is indeed the most convenient 
for our purposes.

 Let us first see how this action comes about from first 
principles. As discussed in section \ref{metricpost}, following Will's book\cite{Willbook} one can argue 
that the EEP can only be satisfied if there exists some metric and the matter 
fields are coupled to it not necessarily minimally but through a 
non-constant scalar, {\em i.e.}, they can be coupled to a 
quantity $\phi g_{\mu\nu}$, where $\phi$ is some scalar. However, this coupling 
should be universal in the sense that all fields should couple 
to $\phi$ in the same way \footnote{This is not the case in 
supergravity and string theories, in which gravivector and 
graviscalar fields can couple differently 
to particles with different quark content\cite{SherkGasperini}.}. So the most general form of the 
matter action will have a dependence on $\phi g_{\mu\nu}$. Of 
course, one can always choose to write $\phi$ as 
$e^{2\alpha(\phi)}$, where $\phi$ is a dynamical field.

Now, the rest of the action should depend on $\phi$, the metric 
and their derivatives. No real principle leads directly to the 
action above. However, one could impose that the resulting field 
equations should be of second order both in the metric and the 
scalar and utilize diffeomorphism invariance arguments to arrive 
to this action. Then, (\ref{action}) is the most general scalar-tensor action that one can write, once no fields  other than $\phi$ and the metric 
are considered, and no other couplings than a non-minimal 
coupling of the scalar to the curvature is allowed.

We now return to the role of the four yet 
to be defined functions $A(\phi)$, $B(\phi)$, $V(\phi)$, $\alpha(\phi)$ and examine whether 
there are redundancies. 
As already said, the action (\ref{action}) describes a {\em 
class} of theories, not a 
single theory. Specifying some of the four functions will pin 
down a specific theory within that class. However, one can 
already see that this action is formally invariant under arbitrary 
conformal transformations  
$\tilde{g}_{\mu\nu}=\Omega^2(\phi) g_{\mu\nu}$. In fact, it can be recast in its initial form by simply redefining the undetermined functions $A(\phi)$, $B(\phi)$, $V(\phi)$, $\alpha(\phi)$ after the conformal transformation. This implies 
that  one can set any one of the functions $A(\phi)$, $B(\phi)$, 
$V(\phi)$ and $e^{2\alpha(\phi)}$ to a (non-vanishing) constant 
through a suitable choice of $\Omega(\phi)$. Additionally, the 
scalar field $\phi$ can be redefined conveniently, in order to 
set yet  another of these functions   to a constant. Therefore, 
we conclude that setting two of these  functions to a constant 
(or just unity) is  
merely a choice of representation and has nothing to do with the 
theory. Actually, it does not even select  a theory within the 
class.

This has a precise physical meaning: it demonstrates our ability 
to choose our clocks and rods at will\cite{Dicke}. One could 
decide not to allow that in a theory  (irrespectively of how 
natural that would 
be). Therefore, it constitutes a very basic physical assumption or even an axiom.

Let us now turn our attention to the matter fields 
$\psi^{(m)}$: the way we have written the action implies that we 
have already chosen a representation for them. However, it 
should be clear that we could always redefine the matter fields 
at will. For example one could set $\tilde{\psi}=\Omega^s  
\psi^{(m)}$ where $s$ is a conveniently selected conformal  
weight\cite{Wald} so that, after a  conformal 
transformation, the matter action will be \begin{equation} 
S^{(m)}=S^{(m)}\left[\tilde{g}_{\mu\nu},\tilde{\psi}\right] \;. 
\end{equation} 
The tilde is used is order to distinguish the physical variable in the two representations.
We can now make use of the previously discussed freedom to fix two of the four functions of the field at will and set $A=B=1$.  The action (\ref{action}) will then take the same form as that of general relativity with a scalar field minimally coupled to gravity. 

However this theory is not general  relativity since 
now $\tilde{\psi}=\tilde{\psi}(\phi)$ which essentially means 
that we have allowed the masses of elementary particles 
and the coupling constants to vary with $\phi$ and consequently 
with their location in spacetime. From a physical perspective, this is translated into 
our  ability to choose whether it will be our clocks and rods 
that are unchanged in time and space or instead the outcome of our measurements\cite{Dicke} (which, remember, are always dimensionless 
constants or dimensionless ratios, as, even the measurement of a dimensional quantity such as {\em e.g.},~a mass, is nothing more than a comparison with a fixed unit of the same dimensions). We will return to this issue again in section \ref{vacuum}.

To  summarize, we can practically choose two of the four 
functions in the action 
(\ref{action}) without specifying the 
theory. In addition, we can fix even a third function at the 
expense of allowing the matter fields $\psi^{(m)}$ to depend 
explicitly on $\phi$, which leads to varying fundamental units\cite{Dicke}. Once any of these two options is chosen, the 
representation is completely fixed  and any further fixing of 
the remaining function or functions leads to a  specific theory within the 
class. On the other hand, by choosing  any two functions and 
allowing for redefinitions of the metric  and the scalar field, 
it is possible to fully specify the theory  and still leave the 
representation completely arbitrary. 

However, it is now obvious 
that each representation might display different characteristics 
of the theory and care should be taken in order not to 
be  misled into representation-biased conclusions, exactly as it 
happens in different coordinate systems. This highlights the 
importance of distinguishing between different {\em theories} 
and different {\em representations}.

This situation is very similar to a gauge theory in which one 
must be careful to derive only gauge-independent results. Every gauge is 
an admissible ``representation'' of the theory, but only 
gauge-invariant quantities should be computed for comparison 
with experiment. In the case of scalar-tensor gravity however, 
it is not clear what a ``gauge'' is and how to identify the  
analog of ``gauge-independent'' quantities.

\subsection{Alternative theories and alternative representations: Jordan and Einstein frames}
\label{JE}

Let us now go one step further and pick up specific 
scalar-tensor theories. With $\psi^{(m)}$ representing the matter fields and by choosing 
$\alpha=0$ and $A(\phi)=\phi$ we fully fix the representation. Let us now 
suppose that all other functions are known. The action takes the 
form 
\begin{equation} 
\label{rep1}
S=S^{(g)}+S^{(m)}\left[g_{\mu\nu},\psi^{(m)}\right]\;,
\end{equation} 
where 
\begin{equation}
 S^{(g)}=\int 
d^{4}x\;\sqrt{-g}\left[\frac{\phi}{16\pi G}R 
- \frac{B(\phi)}{2}g^{\mu\nu}\nabla_{\mu}\phi\nabla_{\nu} 
\phi-V(\phi)\right] \;,
\end{equation} 
and it is apparent that $T_{\mu\nu}\equiv  -(2/\sqrt{-g})\delta 
S^{(m)}/\delta g^{\mu\nu}$ is divergence-free 
with respect to the metric $g_{\mu\nu}$ and, therefore, the 
metric postulates are satisfied.

Now let us take instead a representation where $A=B=1$. Then the action (\ref{action}) takes the form 
\begin{equation} 
\label{rep2}
S=S^{(g)}+S^{(m)}\left[e^{2\tilde{\alpha}(\phi)}\tilde{g}_{\mu\nu}, 
\psi^{(m)}\right]\;,
\end{equation} 
with
\begin{equation} S^{(g)}=\int  
d^{4}x\;\sqrt{-\tilde{g}}\left[\frac{1}{16\pi G}\tilde{R} 
-\frac{1}{2}\tilde{g}^{\mu\nu}\tilde{\nabla}_{\mu} 
\phi\tilde{\nabla}_{\nu}\phi-\tilde{V}(\phi)\right] .
\end{equation} 
As we have argued, for any (non-pathological) choice of $B$ and 
$V$ in the action (\ref{rep1}) there exist some conformal factor 
$\Omega(\phi)$    relating $g_{\mu\nu}$ and $\tilde{g}_{\mu\nu}$ 
and some suitable  redefinition of the scalar $\phi$ to the 
scalar $\tilde{\phi} $, which brings action (\ref{rep1}) to the 
form of action (\ref{rep2}), therefore relating $B$ and $V$ 
with 
$\tilde{V}$ and $\tilde{\alpha}$. Actions (\ref{rep1}) and 
(\ref{rep2}) are just different representations of the same 
theory after all, assuming that $B$ and $V$ or $\tilde{V}$ and 
$\tilde{\alpha}$ are known. 

 According to most frequently used terminology, the first 
representation is called the {\em 
Jordan frame} and 
the second the {\em Einstein frame} and the way we have just 
introduced them should  make it crystal clear that they are just 
 alternative, but physically equivalent, representations, of the 
same theory. (Furthermore,  infinitely many conformal frames 
are  possible, corresponding to the freedom of choosing the 
conformal factor.)

Let  us note however that if one defines the stress-energy 
tensor  in the Einstein frame as $\tilde{T}_{\mu\nu}\equiv 
-(2/ \sqrt{-\tilde{g}})\delta  S^{(m)}/\delta 
 \tilde{g}^{\mu\nu}$ one can show that  it is {\em not} 
 divergence-free with  respect to the Levi-Civita connection of 
 the metric $\tilde{g}_{\mu\nu}$. In fact the transformation 
 property of the matter stress-energy tensor 
under the conformal transformation $g_{\mu\nu}\rightarrow 
\tilde{g}_{\mu\nu}=\Omega^2 \, g_{\mu\nu} $ is 
$\tilde{T}_{\mu\nu}=\Omega^s \, T_{\mu\nu}$, where the 
appropriate conformal weight in four 
spacetime dimensions is $s=-6$ \cite{Wald}. Then, the 
Jordan frame covariant 
conservation equation $\nabla^{\beta}T_{\alpha\beta}=0$ is 
mapped into the Einstein frame equation
\begin{equation}\label{non-conservation}
\tilde{\nabla}_{\alpha}\tilde{T}^{\alpha\beta}=-\tilde{T}\, 
\, \frac{ \tilde{g}^{\alpha\beta} \tilde{\nabla}_{\alpha} 
\Omega}{\Omega} \;,
\end{equation}
which highlights the fact that the Einstein frame  energy-momentum  tensor of matter is not covariantly conserved, unless it describes conformally invariant matter with vanishing trace $T$, which of course is not the general case.

In summary, we see that while the actions (\ref{rep1}) and (\ref{rep2}) are just different  representations of the same theory, the metric postulates and the EEP are 
obviously satisfied in terms of the variables of the Jordan frame, whereas, at least  
judging naively from eq.~(\ref{non-conservation}), one could be led to  the conclusion that the the EEP is not satisfied by  the variables of the Einstein frame representation. However this is obviously paradoxical as we have seen that the general form of the scalar-tensor action (\ref{eq:3.19}) can be derived from the EEP.

The point is that an experiment is not sensitive to the  representation, and hence in the case of the action (\ref{eq:3.19}) it will not show any violation of the EEP. The EEP will {\em not} be violated in {\em any} chosen representation of the theory. A 
common  misconception is that people speak about violation of the EEP or the WEP in 
the Einstein frame simply implying that $\tilde{g}_{\mu\nu}$ is not the 
metric whose geodesics coincide with free fall trajectories. 
Even though this is correct, it does not imply a violation of 
the WEP or the EEP simply because all that these principles 
require is that there exist {\em some} metric whose geodesics 
coincide with free fall trajectories, and indeed we have one, namely $g_{\mu\nu}$, the metric tensor of the Jordan frame. Whether or not one chooses to represent the theory with respect to this metric is simply irrelevant.

To go one step further let us study free fall trajectories in  the Einstein frame. By 
considering a dust fluid with stress-energy tensor 
$\tilde{T}_{\alpha\beta} 
= \tilde{\rho}\,\tilde{u}_{\alpha} \tilde{u}_{\beta}$, eq.~(\ref{non-conservation}) becomes
\begin{equation}
\tilde{\nabla}_{\alpha}\left( \tilde{\rho}\,\tilde{u}^{\alpha}
\tilde{u}^{\beta} \right)=
\tilde{\rho} \,
\frac{\tilde{g}^{\alpha\beta}\,\tilde{\nabla}_{\alpha} \Omega}{\Omega} \;.
\end{equation}
By projecting this equation onto the 3-space orthogonal to 
$\tilde{u}^{\mu}$ by means of the the operator 
$\tilde{h}^{\mu}_{\nu}$ defined by $ 
\tilde{g}_{\mu\nu}=-\tilde{u}_{\mu}\tilde{u}_{\nu} + 
\tilde{h}_{\mu\nu}$ and satisfying  
$\tilde{h}^{\alpha}_{\beta}\,\tilde{u}^{\beta}=0$, one obtains
\begin{equation}\label{correctedgeodesic}
\tilde{a}^{\gamma}\equiv 
 \tilde{h}^{\gamma}_{\beta}\tilde{u}^{\alpha} 
\tilde{\nabla}_{\alpha}\tilde{u}^{\beta}=\tilde{h}^{\gamma\alpha}
\frac{\partial_{\alpha} \Omega (\phi) }{\Omega (\phi) } \;.
\end{equation}
 The term on the right hand side of eq.~(\ref{correctedgeodesic}), which would have been zero if the latter was the standard  geodesic equation, can be 
seen as due to 
the gradient of the scalar field $\phi$, or as due to the 
variation of the particle mass $\tilde{m}=\Omega^{-1} \, m $ 
along its trajectory, or as  due to the variation of the 
Einstein frame unit of mass $\tilde{m}_u =\Omega^{-1}\, m_u $ 
(where $m_u$ is the constant unit of mass in the Jordan frame) 
with the spacetime point --- see Faraoni and Nadeau\cite{VFSN} for an 
extensive 
discussion.

Massive particles in the Einstein frame are {\em 
always} subject to a force proportional to $\nabla^{\mu}\phi$, 
hence there are no massive test particles in this representation 
of the theory. From this perspective, the formulation of EEP ``(massive) test 
particles 
follow (timelike) geodesics'' is not satisfied nor violated: it 
is simply empty. Clearly, {\em the popular formulation of the EEP in terms of the metric postulates is representation-dependent}.

The metric $g_{\mu\nu}$ has in this sense a distinguished status 
with respect to any other conformal metric, such as 
$\tilde{g}_{\mu\nu}$. However, it is a matter of taste 
and sometimes misleading to call a representation physical or 
not. The fact that it is better highlighted in the Jordan frame 
that the theory under discussion satisfies the EEP does not make 
this frame preferable, in the same sense that the Local Lorentz 
coordinate frame is not a preferred one. The Einstein frame is 
much more suitable for other applications, {\em e.g.}, finding 
new exact solutions by using mappings from the Einstein 
conformal frame, 
or the computation of the spectrum of density perturbations 
during inflation in the early universe.

Let us now concentrate on the ambiguities related to the 
metric postulates mentioned in sections \ref{set} and \ref{matgeom}. One should be already convinced that these postulates 
should be generalized to include the phrase ``there exists a 
representation in which'' (Thorne and Will themselves comment 
on the dependence of their postulates on the representation\cite{ThorneWill}). But 
apart from that, there are additional problems. For example, in 
the  Jordan frame $\phi$  couples explicitly to the Ricci 
scalar. One could,  therefore, say that $\phi$ is a 
gravitational field and not a 
matter field. In the Einstein frame, however, $\phi$ is not 
coupled  to the Ricci scalar---it is actually minimally coupled 
to gravity  and non-minimally coupled to matter. Can then one 
consider it a  matter field? If this is the case then maybe one should 
define  the stress-energy tensor differently than before and include the 
$\phi$  terms in the action, {\em i.e.}, define 
\begin{eqnarray} 
\label{rep3}
\bar{S}^{(m)}& = & \int  d^{4}x\; 
\sqrt{-\tilde{g}}\left[-\frac{1}{2} 
\tilde{g}^{\mu\nu}\tilde{\nabla}_{\mu}\tilde{\phi}\tilde{\nabla}_{\nu}\tilde{\phi}-\tilde{V}(\tilde{\phi})\right] \nonumber \\
&& \nonumber \\
&+&  S^{(m)}\left[e^{2\tilde{\alpha}(\tilde{\phi})}\tilde{g}_{\mu\nu},\psi^{(m)}\right] 
\end{eqnarray} 
and 
\be
\label{setbar}
\bar{T}_{\mu\nu}\equiv -(2/\sqrt{-\tilde{g}})\delta 
\bar{S}^{(m)}/\delta \tilde{g}^{\mu\nu}.
\ee

In this case though, $\bar{T}_{\mu\nu}$ will indeed be 
divergence-free with respect to $\tilde{g}_{\mu\nu}$! The 
easiest way to see that is to consider the field equations that  
one  derives from the action (\ref{rep2}) through a variation 
with  respect to $\tilde{g}_{\mu\nu}$ and once the redefinitions 
of  eqs.~(\ref{rep2}) and (\ref{setbar}) are taken into account. 
These are
\be
\tilde{G}_{\mu\nu}=\kappa\bar{T}_{\mu\nu},
\ee
where $\tilde{G}_{\mu\nu}$ is the Einstein tensor of the metric $\tilde{g}_{\mu\nu}$. The contracted Bianchi identity $\tilde{\nabla}_\mu \tilde{G}^{\mu\nu}=0$ directly implies that 
$\tilde{\nabla}_\mu \bar{T}^{\mu\nu}=0$. 

Does that solve 
the problem, 
and the fact that it was not apparent that the EEP is not 
violated in the Einstein frame was just due to a wrong choice of 
the stress-energy tensor?  Unfortunately, this is not the case. First of all $\tilde{g}^{\mu\nu}$ is still 
not the metric whose geodesics coincide with free fall 
trajectories, as it has been shown. Secondly, $\bar{T}_{\mu\nu}$ has the following form
\be
\bar{T}_{\mu\nu}=\tilde{\nabla}_{\mu}\tilde{\phi}\tilde{\nabla}_{\nu}\tilde{\phi}-\frac{1}{2}\tilde{g}_{\mu\nu}\tilde{\nabla}^{\sigma}\tilde{\phi}\tilde{\nabla}_{\sigma}\tilde{\phi}-\tilde{g}_{\mu\nu}\, \tilde{V}(\tilde{\phi})+\tilde{T}_{\mu\nu},
\ee
with $\tilde{T}_{\mu\nu}$ depending on $\tilde{\phi}$ as well, and it will not reduce to the 
 special-relativistic stress-energy tensor for the matter field 
$\psi^{(m)}$ if $\tilde{g}_{\mu\nu}$ is 
 taken to be flat. The same is true for the action 
$\bar{S}^{(m)}$. Both of  these features are due to the fact 
that $\bar{T}_{\mu\nu}$  includes a non-minimal coupling between 
the matter fields $ \psi^{(m)}$ and the scalar $\phi$. Actually, 
setting  $\tilde{g}_{\mu\nu}$ equal to the Minkowski metric does 
not correspond to 
choosing the Local Lorentz frame: that would be the one in which 
$g_{\mu\nu}$ is flat to second order (see section \ref{metricpost}).

The moral is that one can find quantities that indeed formally 
satisfy the metric postulates but these quantities are not 
necessarily physically meaningful. There are great ambiguities 
as mentioned before, in defining the stress-energy tensor or in
judging whether a field is gravitational or just a matter field 
that practically makes the metric postulates useless outside of 
a 
specific representation (and how does one know, in general, when 
given an action, if it is in this representation, {\em  
i.e.,}~if 
the quantities of this representation are the ones to be used directly to check the validity of the metric postulates or a representation change is due before doing so?).

\subsection{Matter or geometry? An ambiguity}
\label{vacuum}

We already saw that treating $\phi$ as a matter field merely because it is minimally coupled to gravity and including it in the stress-energy tensor did not help in clarifying the ambiguities of the metric postulates. Since, however, this did not answer the question of whether a field should be considered of gravitational (``geometric'') or 
of non-gravitational (``matter'') nature, let us try to get some further insight.

Consider again, as an example, scalar-tensor 
gravity. Choosing, $A(\phi)=8\pi\,G\,\phi$ and $\alpha$ to be a constant, the action (\ref{action}) can be written as 
\begin{eqnarray}
S&=&\int d^4 x \sqrt{-g} \left[ \frac{\phi R}{2} 
-\frac{B(\phi)}{2}\, g^{\mu\nu} \nabla_{\mu}\phi 
\nabla_{\nu}\phi -V(\phi) \right. \nonumber \\
&& \nonumber \\
&& \left. +\alpha_{\psi}{\cal L}^{(\psi)}\left( 
g_{\mu\nu}, \psi^{(m)} \right) \right]\;,\label{Jframeaction}
\end{eqnarray}
where $\alpha_{\psi}$ is the coupling constant between gravity 
and the specific matter field $\psi^{(m)}$ described by the 
Lagrangian density ${\cal L}^{(\psi)}$. This representation is 
the Jordan frame and it is no different than that of the  action 
(\ref{rep1}), apart from the fact that we have not specified  
the value of the coupling constant $\alpha_{\psi}$ to be $1$.

It is common practice to 
say that the Brans-Dicke scalar field $\phi$ is  gravitational, 
{\em i.e.}, it describes gravity together with the metric 
$g_{\mu\nu} 
$ \cite{Willbook,Wagoner,Dicke,BD}. Indeed, $1/\phi$ 
plays the role of a (variable) gravitational coupling. However, 
this interpretation only holds in the Jordan frame. As, 
discussed  earlier, the 
conformal transformation to the Einstein frame $g_{\alpha\beta} 
\rightarrow \tilde{g}_{\alpha\beta}=\Omega^2 \, g_{\alpha\beta}$  
with $\Omega=\sqrt{G \phi}$, together with the scalar field 
redefinition
\begin{equation}\label{scalarredefinition}
d\tilde{\phi}=\sqrt{ \frac{ 2\omega(\phi)+3}{16\pi G}}\, 
\frac{d\phi}{\phi}
\end{equation}
casts the action into the form
\begin{equation}\label{Eframeaction}
S=\int d^4 x \sqrt{-\tilde{g}} \left[ \frac{\tilde{R} }{2} 
-\frac{1}{2}\, \tilde{g}^{\mu\nu} 
\tilde{\nabla}_{\mu}\tilde{\phi} 
\tilde{\nabla}_{\nu}\tilde{\phi} 
-\tilde{V}\left( \tilde{\phi} \right)+\tilde{\alpha}_{\psi}{\cal 
L}^{(\psi)}\right] \;,
\end{equation}
where 
\begin{equation}
\tilde{V}\left( \tilde{\phi} \right)=\frac{ V\left[ \phi \left( 
\tilde{\phi} \right) \right] }{ \phi^2\left( \tilde{\phi} 
\right)}
\end{equation} 
and 
\begin{equation} 
\tilde{\alpha}_{\psi}\left( \tilde{\phi} 
\right)=\frac{ 
\alpha_{\psi} }{ \phi^2\left( \tilde{\phi} 
\right)} \;.
\end{equation} 
The ``new'' scalar field $\tilde{\phi}$ is now minimally coupled 
to the Einstein frame Ricci scalar $\tilde{R}$ and has canonical 
kinetic energy: {\em a priori}, nothing forbids to interpret 
$\tilde{\phi}$ as being a ``matter field''. The only memory of 
its gravitational origin as seen from the Jordan frame is in 
the fact that  now $\tilde{\phi}$ couples non-minimally to 
matter, as described by the varying coupling 
$\tilde{\alpha}_{\psi}(\tilde{\phi})$ \footnote{This point was also stressed by Weinstein\cite{Wein}.}.  But, by itself, this 
coupling 
only describes an interaction between $\tilde{\phi}$ and the 
``true'' matter field $\psi^{(m)}$. One could, for example, take 
$\psi^{(m)}$ as the Maxwell field and consider an axion field 
that couples explicitly to it, obtaining an action similar to 
(\ref{Eframeaction}) and being unable to discriminate between 
this axion and  a putative ``geometrical'' field on the base of 
its non-minimal coupling. Even worse, this ``anomalous'' 
coupling of $\tilde{\phi}$ to matter is lost if one considers 
only the gravitational sector of the theory by dropping ${\cal L}^{(\psi)}$ from the discussion.  
This is the situation, for example,  if the scalar 
$\tilde{\phi}$ is supposed to dominate the dynamics of an early, 
inflationary, universe or of a late, quintessence-dominated, 
universe.

More generally, the distinction between gravity and matter 
(``gravitational'' versus ``non-gravitational'') becomes blurred 
in any change of representation involving a conformal 
transformation of the metric 
$g_{\mu\nu}\rightarrow  \tilde{g}_{\mu\nu}=\Omega^2 \, 
g_{\mu\nu} $. The transformation property of the Ricci tensor is 
\cite{Wald,Synge}
\bea\label{Riccitransform}
\tilde{R}_{\alpha\beta}&=& R_{\alpha\beta}
-2\nabla_{\alpha}\nabla_{\beta} \left( \ln \Omega \right) 
-g_{\alpha\beta} g^{\gamma\delta} \nabla_{\gamma}\nabla_{\delta} 
\left( \ln \Omega \right) \nonumber\\ & &
+2 \left( \nabla_{\alpha}\ln \Omega \right)
\left( \nabla_{\beta}\ln \Omega \right)
-2g_{\alpha\beta} g^{\gamma\delta} 
\left( \nabla_{\gamma}\ln \Omega \right)
\left( \nabla_{\delta}\ln \Omega \right) \;.
\eea
A vacuum solution ({\em i.e.}, one with $R_{\alpha\beta}=0$) in 
the 
Jordan frame is mapped into a non-vacuum solution 
($\tilde{R}_{\alpha\beta}\neq 0$) in the Einstein frame by the 
conformal transformation. The conformal factor $\Omega$, a  
purely ``geometrical'' field in the Jordan frame is now playing 
the role of a form of ``matter'' in the Einstein frame. 

A possible way of keeping track of the gravitational nature of 
$\Omega$  is by remembering that the Einstein frame units of 
time, length, and mass are not constant but scale according to 
$\tilde{t}_u=\Omega\, t_u$, 
$\tilde{l}_u=\Omega\, l_u$, and 
$\tilde{m}_u=\Omega^{-1} \, m_u$, respectively (where
$t_u, l_u$, and $m_u$ are the corresponding constant units in 
the Jordan frame)\cite{Dicke}. However, one would not know this 
prescription by looking only at the Einstein frame action 
(\ref{Eframeaction}) unless the prescription for the units is 
made part of the theory ({\em i.e.,}~carrying extra information 
with respect to the one given by the  action!). In practice, even when the action (\ref{Eframeaction}) 
is explicitly obtained from the Jordan frame representation, the 
variation of units with $\Omega$ (and therefore with the 
spacetime point) is most often forgotten in the literature\cite{VFSN} hence leading to the study of a different theory with respect to that expressed by the action (\ref{Jframeaction}).

Going back to the distinction between material and 
gravitational fields, an 
alternative possibility to distinguish between ``matter'' and 
``geometry'' would seem to arise by labeling  as ``matter 
fields'' only those described by a stress-energy tensor that 
satisfies some energy condition. In fact, a  conformally 
transformed field that originates from Jordan frame geometry 
does not, in general, satisfy any energy condition. The 
``effective stress-energy tensor'' of the field $\Omega$ derived 
from eq.~(\ref{Riccitransform}) does not have the canonical 
structure quadratic in the first derivatives of the field but 
contains instead terms that are linear in the second 
derivatives. Because of this structure, the stress-energy tensor $\Omega$ violates the 
energy conditions. While it would seem that labelling as ``matter 
fields'' those that satisfy the weak or null energy condition 
could eliminate the ambiguity, this is not the case. As we have previously seen,  one can always redefine the scalar field in such a way that it is minimally coupled to 
gravity and has canonical kinetic energy (this is precisely the 
purpose of the field redefinition~(\ref{scalarredefinition})). 
Keeping track of the 
transformation of units in what amounts to a full specification 
of the representation adopted  (action plus information on 
how the units scale with the scalar field) could help making the 
property of satisfying energy conditions frame-invariant, but at the cost of extra ``structure'' in defining a given theory.

As  a conclusion, the concept of vacuum versus non-vacuum, or of 
``matter field'' versus ``gravitational field'' is 
representation-dependent. One might be prepared to accept {\em a priori} and without any real physical justification that one representation should be chosen in which the fields are to be characterized as gravitational or non-gravitational and might be willing to carry this extra ``baggage" in any other representation in the way described above. Even so, a solution to the problem which would be as tidy  as one would  like, is still not provided.

These considerations, as well those discussed at the end of section \ref{JE}, elucidate a more general point: it is not 
only the mathematical formalism associated with a theory that is 
important, but the theory must also include a set of rules to 
interpret physically the mathematical laws. As an example 
from the classical mechanics of point particles, consider two 
coupled harmonic oscillators described by the Lagrangian
\begin{equation}\label{Lq}
L=\frac{\dot{q}_1^2}{2}+ \frac{\dot{q}_2^2}{2}-
\frac{q_1^2}{2}-\frac{q_2^2}{2}+\alpha \, q_1 q_2 \;.
\end{equation}
A different representation of this physical system is obtained 
 by using normal coordinates $ Q_1\left( q_1, q_2 \right) , 
Q_2\left( q_1, q_2 \right) $, in terms of which the Lagrangian 
(\ref{Lq}) becomes
\begin{equation}\label{LQ}
L=\frac{\dot{Q}_1^2}{2}+ \frac{\dot{Q}_2^2}{2}-
\frac{Q_1^2}{2}-\frac{Q_2^2}{2} \;.
\end{equation}
Taken at face value, this Lagrangian describes a different 
physical system, but we know that the mathematical expression 
(\ref{LQ}) is not all there is to the theory: the {\em 
interpretation} of $q_1$ and $q_2$ as the degrees of freedom of 
the two original oscillators prevents viewing $Q_1$ and 
$Q_2$  as the physically measurable quantities. In addition to the equations of 
motion, a set of interpretive rules constitutes a fundamental 
part of a theory. Without such rules it is not only impossible to connect the results derived through the mathematical formalism to a physical phenomenology but one would not even be able to distinguish alternative theories from alternative representations of the same theory.  Note however, that once the interpretative rules are assigned to the variables in a given representation they do allow to predict the outcome of experiments in  any other given representation of the theory (if consistently applied), hence assuring the physical equivalence of the possible representations.

While the above comments hold in general for any physical theory,
it must however be stressed that gravitation theories are one of those cases in which the problem is more acute. 
In fact, while the physical interpretation of the variables is clear in simple systems, such as the example of the two coupled oscillators discussed above, the physical content of  complex theories (like quantum mechanics or gravitation theories) is far less intuitive. Indeed, for what regards gravity, what we actually  
know more about is the phenomenology of the system instead of the system itself. Therefore, it is often difficult, or even arbitrary, to formulate explicit interpretive rules, which should nevertheless be provided in order to completely specify the theory.

\section{$f(R)$ gravity}
\label{fR}

To highlight even more the ambiguity of whether a field is a 
gravitational or matter field, as well as to demostrate how the problems discussed here can actually go beyond representations that just involve conformal redefinitions of the metric, let us utilise one further 
example: that of $f(R)$ gravity\cite{buh} (see also \cite{thesis} for a recent review and references). 
These gravity  theories
have received considerable attention lately, as they lead to 
interesting cosmological phenomenology which may be able to 
account for dark energy\cite{fofrcosm} \footnote{Several 
concerns have been expressed  about the viability of these 
theories, and in many cases  there is still open debate\cite{viability}. However,  viability is irrelevant here as we 
are using these theories as  examples to make a completely 
different point.}. The action of the theory is
\be
\label{metaction}
S_{met}=\frac{1}{2\kappa}\int d^4 x \sqrt{-g} \, f(R) +S_M(g_{\mu\nu},\psi).
\ee
Variation with respect to the metric gives\cite{buh}
\be
\label{metf}
f'(R)R_{\mu\nu}-\frac{1}{2}f(R)g_{\mu\nu}-\nabla_\mu\nabla_\nu f'(R)+g_{\mu\nu}\Box f'=\kappa T_{\mu\nu},
\ee
where a prime denotes differentiation with respect to the argument. These are fourth order partial differential equations for the metric.

One could also choose to consider the connection $\Gamma^{\lambda}_{\phantom{a}\mu\nu}$ as an independent quantity and construct the Riemann tensor and the Ricci tensor accordingly. The Ricci tensor, ${\cal R}_{\mu\nu}$, is constructed using only the connections and the scalar curvature
${\cal R}$ is then defined to be the contraction of
${\cal R}_{\mu\nu}$ with the metric, {\em i.e.},  
${\cal R}\equiv g^{\mu\nu}{\cal R}_{\mu\nu}$.
The action will then be formally the same but $R$ will be 
replaced with ${\cal R}$, {\em  i.e.,}
\be
\label{palaction}
S_{pal}=\frac{1}{2\kappa}\int d^4 x \sqrt{-g} f({\cal R}) +S_M(g_{\mu\nu}, \psi).
\ee
Notice that the matter action is chosen not to depend on the 
independent connection. Were the matter action allowed 
to depend on the independent connection, the resulting theory 
would be a metric-affine theory of gravity\cite{metaff1,metaff2}.
 Independent (Palatini) variations with 
respect to the metric and the connection lead to the two field 
equations 
\bea
\label{field1sym}
& &f'({\cal R}) {\cal R}_{(\mu\nu)}-\frac{1}{2}f({\cal 
R})g_{\mu\nu}=\kappa T_{\mu\nu},\\
\label{field2sym}
& &\stackrel{\Gamma}{\nabla}_\lambda\left(\sqrt{-g}f'({\cal 
R})g^{\mu\nu}\right)=0,
\eea
one for the metric and one for the independent  
connection. $\stackrel{\Gamma}{\nabla}_{\mu}$ denotes the covariant 
derivative defined with the independent connection 
$\Gamma^{\lambda}_{\phantom{a}\mu\nu}$. Theories described by 
the action (\ref{palaction}) are called $f(R)$ theories of 
gravity in the Palatini formalism or simply Palatini $f(R)$ 
theories of gravity.

Now, consider the action (\ref{metaction}) of metric $f(R)$ 
gravity. One can introduce a new field $\chi$ and write
a dynamically equivalent action\cite{STequivalence}:
\be
\label{metactionH}
 S_{met}=\frac{1}{2\kappa}\int d^4 x \sqrt{-g} 
\left[f(\chi)+f'(\chi)(R-\chi)\right] +S_M(g_{\mu\nu},\psi).
\ee
Variation with respect to $\chi$ leads to the equation $\chi=R$ 
if $f''(\chi)\neq 0$, which reproduces action (\ref{metaction}).
Redefining the field $\chi$ by $\phi=f'(\chi)$ and setting
\be
\label{defV}
V(\phi)=\chi(\phi)\phi-f(\chi(\phi)),
\ee
 the action takes the form
\be
\label{metactionH2}
S_{met}=\frac{1}{2\kappa}\int d^4 x \sqrt{-g} \left[\phi R-V(\phi)\right] +S_M(g_{\mu\nu},\psi).
\ee
This is the action of a Brans--Dicke theory
with Brans--Dicke parameter $\omega_0=0$, or the specific choice of $A=\phi$, $B=0$, $\alpha=0$ when one refers to the action (\ref{action}) (fixing both the theory and the representation). So, metric $f(R)$ theories, as has been observed long ago, are fully equivalent to a class of Brans--Dicke theories with vanishing kinetic term\cite{STequivalence}. 

Similar things can be said for the action (\ref{palaction}) for 
Palatini $f(R)$ gravity. Introducing the scalar field $\chi$ as 
before and  redefining it by using $\phi$, the action takes the 
form:
\be
\label{palactionH2}
S_{pal}=\frac{1}{2\kappa}\int d^4 x \sqrt{-g} \left[\phi {\cal R}-V(\phi)\right] +S_M(g_{\mu\nu}, \psi).
\ee
Even though the gravitational part of this action is formally the same as that of action (\ref{metactionH2}), this action
is not a subcase of action (\ref{action}) as ${\cal R}$ is not 
the Ricci scalar of the metric $g_{\mu\nu}$. However, 
eq.~(\ref{field2sym}) implies that the connections are the Levi-Civita connections of the metric $h_{\mu\nu}=f'({\cal R})g_{\mu\nu}$   \cite{fer}.
Using the  definition of $\phi$ we can write $h_{\mu\nu}=\phi 
g_{\mu\nu}$.  Then we can express ${\cal R}$  in terms of $R$ 
and $\phi$:
\be
{\cal R}=R+\frac{3}{2\phi^2}\nabla_\mu \phi \nabla^\mu \phi-\frac{3}{\phi}\Box \phi.
\ee
Substituting in the action (\ref{palactionH2}) yields
\be
\label{palactionH2d0}
S_{pal}=\frac{1}{2\kappa}\int d^4 x \sqrt{-g} \left[\phi R+\frac{3}{2\phi}\nabla_\mu \phi \,\nabla^\mu \phi-V(\phi)\right] +S_M(g_{\mu\nu}, \psi),
\ee
where we have neglected a total divergence. The matter action has now no dependence on the independent connection $\Gamma^{\lambda}_{\phantom{a}\mu\nu}$. Therefore, this is indeed the action of a Brans--Dicke theory with Brans--Dicke parameter
$\omega_0=-3/2$, or, in terms of the action (\ref{action}), it corresponds to the choice $A=\phi$, $B=-3/2$, $\alpha=0$.

Notice that the general representation used in the 
action~(\ref{action}) is actually not as general as one may 
expect, as we have just shown that theories that are indeed 
described by this action under suitable choices of the 
parameters, can even acquire completely different, 
 non-conformal, representations. One can, in principle, add at will
 auxiliary fields, such as the scalar field $\chi$ used above, 
  in order to change the representation of a theory and 
these fields 
  need not necessarily be scalar fields. Therefore, all of the 
problems described so far in this paper are not specific to 
conformal representations. 
In this $f(R)$ representation the scalar $\phi$ is not even there, so 
how one can decide if it is a gravitational or matter field? For 
the   case of metric $f(R)$ gravity, the scalar field was 
eliminated  without introducing any other field, and the metric 
became the  only field describing gravity. On the 
other hand,  in the Palatini formalism the outcome is even more 
surprising if  one considers that the scalar field was replaced 
with an  independent connection, which, theoretically speaking, 
could have forty degrees of 
freedom assuming that it is symmetric, 
and in practice it has only one!

\section{Einstein-Cartan-Sciama-Kibble theory}
\label{ECSK}

 Our final example is 
Einstein-Cartan-Sciama-Kibble 
theory. In this theory, one starts with a metric and an 
independent connection which is not symmetric but has zero 
non-metricity. We will avoid to present here extensive 
calculations and details. Instead, we address the reader to  
the thorough review of Hehl {\em et al}\cite{hehl} (see also Shapiro\cite{shapirorev} for a review on torsion and its quantum aspects). What we would like to focus 
on is the fact that, as the theory has an independent connection, one    
usually arrives to the field equations through independent 
 variations with respect to the metric and the connection. 
 Additionally, since the matter action depends on both the 
 metric and the connection, its variation will lead to two 
 objects describing the matter fields: the stress-energy tensor 
 $T_{\mu\nu}$, which is the product of the variation of the 
 matter action with respect to the metric as usual, and the 
 hypermomentum $\Delta^\lambda_{\phantom{a}\mu\nu}$ which comes 
out of  the variation of the matter action with respect to the 
independent connection.

 In this theory $T_{\mu\nu}$ is not divergence-free with respect 
to either the covariant derivative defined with the 
Levi-Civita connection, or with respect to the one defined with 
respect to the independent connection. On the other hand, it 
also does not reduce to the special-relativistic stress-energy 
tensor in the suitable limit. However, it can be shown, that a 
suitable, non-trivial, combination of $T_{\mu\nu}$ and 
$\Delta^\lambda_{\phantom{a}\mu\nu}$ can lead to a tensor that 
indeed has the latter            property\cite{hehl}. What is 
more, a third connection can be  defined, which leads to a 
covariant derivative with respect  to which this tensor is 
divergence-free\cite{hehl}! That is sufficient to guarantee 
that 
the EEP is satisfied. Does this make 
Einstein-Cartan a metric theory? And how useful are the metric 
postulates to discuss violations of the EEP if, in order to 
show that they are satisfied, one will have already shown geodesic 
motion or LLI on the way?

\section{Discussion and conclusions}
\label{discussion}

We have attempted to  shed some light on the difference between 
different theories and different representations of the same  
theory, as well as to reveal the important role played by a 
representation in our understanding of a theory. To this end, 
several examples which hopefully highlight this issue have been 
presented. It has  been argued that certain conclusions about a 
theory which may  be drawn in a straightforward manner in one 
representation,  might require serious effort when a different 
representation is  used and vice-versa. Additionally, care 
should be taken as certain representations  may be completely 
inconvenient or even misleading for specific applications.

 It is worth commenting at this point, that the literature is 
 seriously biased towards particular representations. 
 Additionally, this bias is not always a result of the 
 convenience of certain representations in a specific 
application, but many times it is a mere outcome of habit. It is 
common, for instance, to bring alternative theories of gravity 
in a general-relativity-like representation due to its familiar  
 form, even if this might be 
misleading when it  comes to the deeper understanding of the 
theory.

So far, this seemingly inevitable representation-dependent 
 formulation of our gravitational theories has already been the 
cause of several misconceptions. What is more, one can very 
easily recognise a representation bias in the definition of 
commonly used quantities, such as the stress-energy tensor. 
Notions such as vacuum and the possibility of distinguishing 
between gravitational fields and matter fields are also 
representation-dependent. This is often overlooked due to the 
fact that one is very much accustomed to the 
representation-dependent  definition given in the literature. On 
the other hand, representation-free definitions do not exist.

Note, that even though the relevant literature focuses almost 
completely on conformal frames, the problems discussed here are 
not restricted to conformal representations.  Even if 
conformally 
invariant theories were considered, nothing forbids the existence of other non-conformal representations of these 
theories under which the action or the field equations will, of 
course, not be invariant. These might be implying that creating 
conformally invariant theories is not the answer to this issue. 
After all, even though measurable quantities are always 
dimensionless ratios and are, consequently, conformally invariant, 
matter is not generically conformally invariant and, therefore, 
neither can (classical) physics be conformally invariant, at 
least when its laws are written in terms of the fields 
representing this matter.

The issue discussed here seems to have its roots in a more 
fundamental problem: the fact that in order to describe a theory 
in mathematical terms, a non-unique set of variables has to be 
chosen. Such a set will always correspond to just one of the 
possible representations of the theory. Therefore, even though 
{\em abstract statements such as the EEP are 
representation-independent}, attempts to turn such statements 
into {\em quantitative mathematical relations that are of practical 
use, such as the metric postulate, turn out to be severely 
representation-dependent}. Moreover, the EP, although 
representation-independent, appears to be of little practical 
use and this is true even if we confine ourselves to the realm of 
classical physics. The mathematical approach towards an axiomatic formulation mentined earlier\cite{mathapproach} may eventually turn out to be more 
convenient in this regard.

The analogy between a choice of a representation and a 
choice of a coordinate system is practically unavoidable. 
Indeed, consider classical mechanics: one can choose a set of
coordinates in order to write down an action describing 
some system. However, such an action can be written in a 
coordinate invariant way. In classical field theory one has to 
choose a set of fields --- a representation --- in order to 
write down the action. From a certain viewpoint, these fields 
are considered to be generalized coordinates. Therefore, one 
could expect that there should be some 
representation-independent way  to describe the theory. However, 
up to this point no real progress has been made on this issue.

The representation dependence of quantitative statements  acts 
in such a way that, instead of merely selecting for us viable 
theories, they 
actually predispose us to choose theories which, in a specific 
representation, naively appear more physically meaningful than 
others irrespectively of whether this is indeed the case. The 
same problem is bound to appear when one attempts to generalise 
a theory and at the same time is biased towards a specific 
representation, as certain generalisations might falsely appear 
as more ``physical'' than others in this representation. This 
effectively answers the question why most of our current 
theories of gravitation eventually turn out to be just different 
representations of the same theory or a class of theories. 
Scalar-tensor theories and theories which include higher order 
curvature invariants, such as $f(R)$ gravity or fourth order 
gravity, are typical examples.

Even though this discussion might at some level appear to be 
 purely philosophical, the practical implication of representation 
dependence should not be underestimated. For instance, how can 
we formulate theories that relate matter/energy and gravity if we 
do not have a clear distinction between the two, or if we cannot 
even conclude whether such a distinction should be made? Should we then aim to drop any statement based on a sharp separation between matter and gravity sectors?

In conclusion, even though significant progress has been made on 
the front of gravitation theories, one cannot help but notice 
that it is still unclear how to relate principles and 
experiments in order to form simple theoretical 
viability criteria expressed in a mathematical way. Our 
inability to enunciate these criteria, as well as several of our 
very basic definitions, in a representation-invariant way seem to 
have played a crucial role in this lack of progress. However, 
this seems to be a critical obstacle to overcome, if we want to 
go beyond a trial-and error approach when it comes to 
gravitational theories.

\section*{Acknowledgments} 

The authors are grateful to Sebastiano Sonego for a critical reading of this manuscript and numerous valuable suggestions for its improvement and to John Miller for helpful comments regarding the presentation of this material. 
V.F. was supported by a Bishop's 
University 
Research Grant, and by the Natural Sciences and Engineering 
Research Council of Canada (NSERC). T.P.S. wishes to thank Bishop's university for its hospitality during the preparation of this work and acknowledges partial support by the Italian MIUR program ``Fundamental Constituents of the Universe''.

   
\end{document}